\documentstyle[twoside,fleqn,espcrc2,epsf]{article}

\newcommand{\AmS}{{\protect\the\textfont2
  A\kern-.1667em\lower.5ex\hbox{M}\kern-.125emS}}

\newcommand{\bee}{\begin{equation}}
\newcommand{\ee}{\end{equation}}
\newcommand{\beea}{\begin{eqnarray}}
\newcommand{\eea}{\end{eqnarray}}

\newcommand{\PR}[3]{{Phys. Rev.} {\bf #1} {(19#2)}  #3}

\title{Optimizing the Chiral Properties of Lattice Fermions}

\author{Thomas DeGrand, Anna Hasenfratz and
        Tam\'as G. Kov\'acs \\[2mm]
        Department of Physics, 
        University of Colorado\\ Boulder, CO 80309-390, USA}

\begin{document}

\begin{abstract}
We describe a way to optimize the chiral behavior  of 
Wilson-type lattice fermion actions by studying the low energy real 
eigenmodes of the Dirac operator.
We find a candidate action, the clover action with fat links with a
tuned clover term.  The action shows good scaling
behavior at Wilson gauge coupling $\beta=5.7$.
\end{abstract}

\maketitle

There is an increasing accumulation of evidence from lattice simulations
that the (lattice) pure gauge
QCD vacuum is filled with instantons of average radius $\sim 0.3$ fm,
with a density of about 1 fm$^{-4}$, and they are responsible for a
large
part of chiral symmetry breaking in QCD \cite{NEGELE}.

Lattice artifacts associated with instantons also seem to be connected
to
some of the difficulties of numerical simulations with quenched QCD, 
namely, eigenmodes of the Dirac operator
which occur away from zero bare quark mass, and which spoil the
calculation of the fermion propagator, 
the so-called ``exceptional configurations'' \cite{FNALINST,NEW_EXCEPT}.

Given that instantons
are responsible for chiral symmetry breaking, is it possible to optimize
the lattice discretization of a fermion action with respect to its
topological
properties? Our   goal is to shrink the range of bare quark mass
over which  the low lying real eigenmodes of the Dirac operator occur,
on background gauge field configurations
which are typical equilibrium configurations
at some gauge coupling. This is a review of our work on this 
topic \cite{DHK_ape}.

The action we propose is the standard clover action, except that
the gauge connections are replaced by APE-blocked
 links, and the
clover coefficient is tuned to optimize chiral properties.
We take $c=0.45$ and $N=10$ smearing steps,
chosen because of our previous work in instantons \cite{COLOINST}.
This choice of parameters is not unique and might not even be optimal.
 At Wilson gauge coupling $5.7-5.8$, the
best choice  of clover coefficient
is $C=1.2$, and it decreases to the tree-level $C=1$ value
at larger $\beta$.

On the lattice, essentially every continuum property of instantons
 is contaminated by lattice artifacts. 
The Wilson  or clover fermion
action analog of $\gamma \cdot D$, $D_w$, is neither
Hermetian nor antihermetian and its eigenvalues are generally complex.
$D_w$ can also have real eigenvalues. These real eigenvalues usually do
not occur at zero bare quark mass.
Their locations spread across a range of quark mass values.

On smooth, isolated instanton background configurations,
the location of the real eigenmode varies with the size $\rho$ of the
instanton.
For large instantons, the low lying real mode occurs close to zero quark
mass.
Accompanying these near-zero modes are a set of modes which 
are not clustered around $m_0=0$, but around
$-am_0 \simeq O(1)$.  These are ``doubler modes.''  
As one decreases the instanton size, the eigenmodes at $am_0 \simeq 0$
shift towards negative quark masses, approaching the doubler modes. 
This $\rho$-dependent mass shift and the 
location of the doubler modes all depend on the particular choice of
lattice fermion action.
The shift of the low lying real eigenmode for a given instanton size for
the Wilson action is larger than for the clover
action, which has better chiral properties. Nevertheless, for both
actions,
as the instanton size decreases,  sooner or later the low energy
eigenmode
shifts to  large negative quark mass,
approaches the doublers and eventually annihilates with one of the
doublers
- the fermion does not see the
instanton any longer.

On equilibrium background configurations,
eigenmodes receive an overall $\beta$-dependent mass shift
and they are spread out according to the
sizes of the background instantons.
The spread of these low energy modes, if distinguishable from the
doublers, characterizes the amount of explicit chiral symmetry
breaking of the fermionic action. 
On configurations with small lattice spacing the typical instanton is
large in
lattice units. The spread of the low energy modes is small and they are
well separated from the doublers. The explicit chiral symmetry
breaking of the fermionic action is small.
On configurations with large lattice spacing the instantons are small
and the spread
of the low lying modes is large,  overlapping with the
doubler modes.
The hadron spectrum on these configurations might be similar
to the 
continuum spectrum, 
but the physical mechanism behind it is very different from continuum
QCD.

We studied the locations of low energy real eigenmodes in equilibrium
(quenched)
gauge configurations to see whether the doublers separated from the near
zero modes, and if so, to measure the width of the near zero modes.

Our goal is to tune the clover coefficient for good chiral behavior,
i.e. to minimize the spread of the
low lying modes. We could attempt this 
program on the original configurations. However, it is  known that rough
gauge configurations with large clover coefficient are plagued by
exceptional configurations. This
problem is greatly reduced if the links of the configurations are
smoothed by a 
series of APE smearing
steps \cite{NEW_EXCEPT}. We also know \cite{COLOINST}
that an
 action with fat
links is insensitive to  short distance fluctuations, but still knows
about instantons and the
additional long distance behavior of the gauge
field responsible for confinement.

\begin{figure}[h!tb]
\begin{center}
\leavevmode
\epsfxsize=70mm
\epsfbox{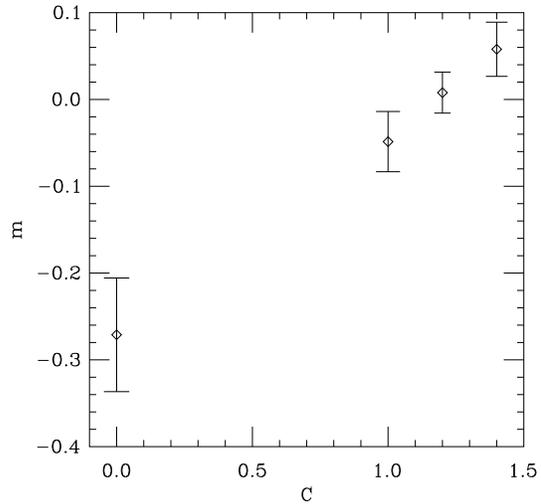}
\end{center}
\vspace{-28pt}
\caption{ The average real eigenmode location as the
function of the clover coefficient at $\beta=5.8$. The error
bars show the spread of the modes. }
\label{fig:averpolea}
\end{figure}

\begin{figure}[h!tb]
\begin{center}
\leavevmode
\epsfxsize=70mm
\epsfbox{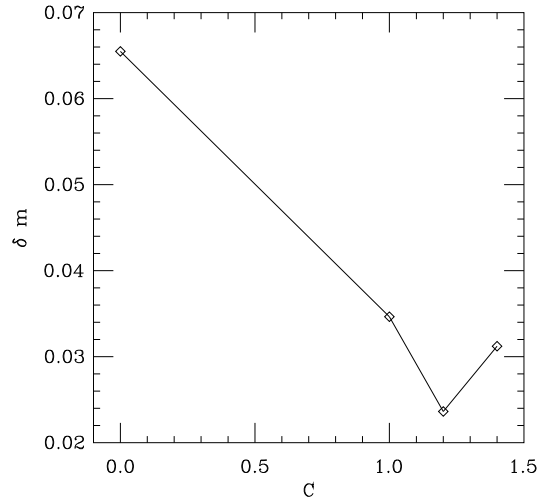}
\end{center}
\vspace{-28pt}
\caption{The spread of the modes from  Fig. 1
as the function of the clover coefficient. }
\label{fig:averpoleb}
\end{figure}

At $\beta=5.8$ the near zero modes and the doublers
of  the fat link clover action are well separated.
In figure \ref{fig:averpolea} we plot the average real eigenmode
location as the function of the clover coefficient. The error bars here
are the spread of the modes in $m$.  
Fig. \ref{fig:averpoleb} shows the spread as a function of the
clover coefficient.
The average eigenmode locations
and the upper end of the error bars in figure \ref{fig:averpolea}
bracket $m_c$, the bare quark mass where the pion becomes massless.
Smearing the link removes most of the additive mass renormalization even
for the Wilson action, reducing $m_c \sim -0.95$ for the thin link action 
to $m_c \sim -0.22$ for our case. 
Adding a clover term to the action further reduces the
additive mass renormalization, and even larger clover terms
induce a positive mass renormalization. At $\beta=5.8$ the 
spread of the eigenmodes is
 minimal for $C\approx 1.2$. This is also the value
where the additive mass renormalization is minimal.
These results indicate that at $\beta=5.8$
 (lattice spacing $a\simeq 0.15$ fm) the low lying  and
doubler modes are well separated and explicit chiral symmetry breaking
is minimized with clover
coefficient $C=1.2$ with our fat link action.

The situation is worse at $\beta=5.7$ (lattice spacing 
$a\simeq 0.2$ fm),
where the low lying  modes are not as well separated.
At $\beta=5.54$ it is worser.
It is apparently not possible to make the lattice spacing
greater than about 0.2 fm, and still retain the continuum-like
description
of chiral symmetry breaking using a clover-like action.

\begin{figure}[h!tb]
\begin{center}
\leavevmode
\epsfxsize=70mm
\epsfbox{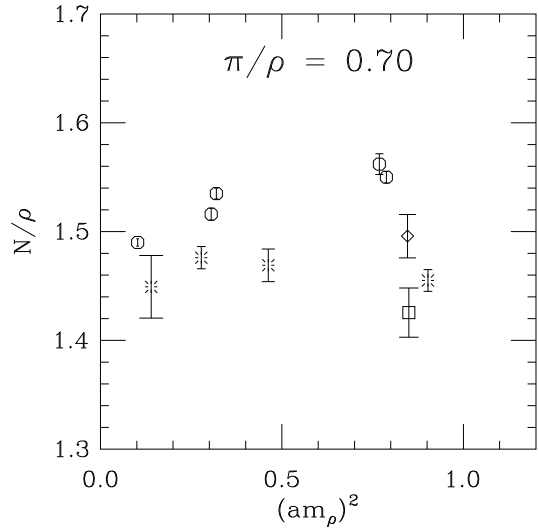}
\end{center}
\vspace{-28pt}
\caption{$m_N/m_\rho$ at $m_\pi/m_\rho=0.7$ for the
tuned action on background configurations of the Wilson
action at $\beta=5.7$ (square), an FP action at $\beta=3.70$
(diamond), the nonperturbative clover action (bursts)
and staggered fermions (octagons).
}
\label{fig:ratio0.7}
\end{figure}

We performed a spectroscopy test at $a=0.2$ fm to test scaling.
The $N/\rho$ mass ratio at $\pi/\rho=0.7$ is shown in Fig. 
\ref{fig:ratio0.7}.
The bursts are from the nonperturbatively improved clover action of
Refs.
\cite{ALPHA} and \cite{SCRI}.
The square shows the C=1.2 action.
The octagons are staggered fermions \cite{MILC}.
The tuned action seems to be no worse than the nonperturbatively
improved action at this lattice spacing. Its dispersion relation
is basically identical to that of the Wilson or clover action.
Because of the large amount of fattening, renormalization factors
for currents are quite close to unity: the Z-factor for the
local current is about 1.025.
We also noticed that the new action seems to be more convergent than 
the thin link clover action: the same biconjugate gradient code needs
about
half as many steps to converge to the same residue as the usual thin
link 
clover action, for the same $\pi/\rho$ mass ratio.
We  could push down to $m_\pi/m_\rho \simeq 0.5$ without
encountering an unacceptable number of exceptional configurations.

We have shown that via a combination of fattening the links and
tuning the magnitude of the clover term,
 it is possible to optimize the chiral behavior of
the clover lattice fermion action.  The example we presented
reduces the spread in the low energy real eigenmodes by about a factor
of three
in units of the squared pion mass, compared to the usual Wilson action.
We fixed the fattening and varied the clover coefficient, but it is
clear that a real optimization would involve varying both factors.

This work was supported by the US Department of Energy.

\end{document}